\documentclass[aps,prl,twocolumn,showpacs,groupedaddress]{revtex4-1}
\usepackage{graphicx}
\begin{document}

\title{Helium at elevated pressures: Quantum liquid with non-static shear rigidity}
\author{D. Bolmatov$^{1}$}
\author{V. V. Brazhkin$^{2}$}
\author{K. Trachenko$^{1,3}$}
\address{$^1$ School of Physics and Astronomy, Queen Mary University of London, Mile End Road, London, E1 4NS, UK}
\address{$^2$ Institute for High Pressure Physics, RAS, 142190, Troitsk, Moscow Region, Russia}
\address{$^3$ South East Physics Network}
\begin{abstract}
The properties of liquid helium have always been a fascinating subject to scientists. The phonon theory of liquids taking into account liquid non-static shear rigidity is employed here for studying internal energy and heat capacity of compressed liquid $^{4}$He. We demonstrate good agreement of calculated and experimental heat capacity of liquid helium at elevated pressures and supercritical temperatures. Unexpectedly helium remains a quantum liquid at elevated pressures for a wide range of temperature supporting both longitudinal and transverse-like phonon excitations. We have found that in the very wide pressure range 5~MPa-500~MPa  liquid helium near melting temperature is both solid-like and quantum.
\end{abstract}

\pacs{67.30.em, 67.30.ef, 65.20.Jk}

\maketitle


\section{Introduction}
This Letter is concerned with the thermodynamic functions of liquid $^{4}$He and their possible relationships to those of its crystalline counterpart. Compared with crystalline and gas phases the statistical description of the structure and thermal properties of liquids remains relatively incomplete. At low pressures, matter commonly exists either as a dense solid or as a dilute vapour. For each of these states there is a model which is a plausible approximation to reality and which constitutes a basis for detailed theoretical extension. These are the ideal periodic arrangements for crystalline structures and the ideal gas proposition, respectively: in the former, emphasis is placed on structural order modified slightly by zero-point or thermal motion of the atoms while (save perhaps for the ultra-cold gases) the latter model describes the thermal motion of the atoms on the basis of random atomic displacements and associated momenta. But as is well known there is a third state of dense matter, the ubiquitous liquid state \cite{gund,coop,elm,bryk,skinner,pet}. This state occurs over a temperature range that separates the regions occupied by the solid and vapour states. However, the problem of formulating a rigorous mathematical description of the molecular motions in liquids has always been regarded as much more difficult than that of the kinetic theory of gases or collective displacements of crystalline solids \cite{widom,bar,chan,ashcroft,cao,chong,skir,anisimov,bryk1,bolm}. Approximation, if not judicious, can lead to a description of either high-density gases or of disordered high-temperature solids. Indeed, at one time considerable effort was devoted to the representation of liquids in these terms, but it is now known that liquids do not have a simple interpolated status between gases and solids, although similarities to the properties of both adjacent phases can certainly be observed.


The ability of liquids and solids to form free bounding surfaces obviously distinguish them from gases. The coefficients of self-diffusion of liquids ( $10^{-5}$ $cm^2$ $s^{-1}$ ) and solids ( $10^{-9}$ $cm^2$ $s^{-1}$ ) are orders of magnitude below those of gases.  And the viscosities of gases and liquids are some thirteen orders of magnitude lower than those of solids, and this we may easily understand in terms of the molecular processes of momentum exchange. In terms of vibrational states liquids differ from solids because they cannot support static shear stress. However, as will be seen below liquids support shear stress at high frequency. Flow in a solid arises primarily from rupturing of bonds and the propagation of dislocations and imperfections. In a liquid flow is characterized by both configurational and kinetic processes, whilst in a gas the flow is understood purely in terms of kinetic transport. In this very limited sense liquids may have a minor partial interpolated status between gases and solids.


The concept of elasticity and viscosity in liquids merit clarifications. Which property dominates, and what values of the associated parameters are assumed, depends on the stress and duration  of application of that stress. If we apply a stress over a very wide spectrum of time, or of frequency, we are able to observe liquid-like properties in solids and solid-like properties in liquids. Frenkel \cite{frenkel} introduced a relaxation time $\tau$ as the average time between two consecutive local structural rearrangements in a liquid. If  $\tau$ is small compared with an observation time it will yield to the process of liquid flow. In this macroscopic hydrodynamic picture, we now have a rather good understanding of most of the fundamental processes operating on such time-scales in liquids, including quantum liquids such as liquid helium.


The investigation of the properties of helium has been one of the most prolific endeavors since its discovery (in 1868) \cite{temp,prigogin,wood}. For most systems the solid phase is the state of lowest energy at one atmosphere. To date liquid helium is the sole exception in this respect: below $1.70^{\circ}$ K liquid helium has a lower free energy than that of solid helium \cite{blondon,simon}. Liquid helium is a quantum liquid at low temperatures and quantum effects play a crucial role. Helium is also of considerable practical significance which is related to the rapidly growing industry surrounding the various applications of superconductivity these often relying on liquid helium as a coolant or refrigerant. Considerable interest is still focused on the highly unusual properties of helium especially at low temperatures and the study of the properties of liquid $^{4}$He continues to be an active area of condensed matter research \cite{,dick,naut,bew,kono,jeff}.


Theory has been long drawn to study the condensed isotopes of helium and their mixtures because these liquids are model many-body systems but with fundamental quantum-statistical differences; they are fertile proving grounds for various quantum many-body formalisms. As Landau emphasized \cite{mandau}, these systems are amenable to theoretical attack because, when studied at relatively low temperatures, they are only weakly excited from their ground states. A description in terms of weakly interacting elementary excitations is then appropriate. The heat capacity of liquid helium has been discussed on just such a basis of elementary excitations and to describe it below the critical temperature Landau suggested two classes of elementary excitations, phonons or quanta of longitudinal compressional waves and rotons \cite{mandau}. The latter are still not completely well understood. The consideration of non-static shear phonon contributions to the energy spectrum of liquid helium has been largely ignored simply because it was not clear whether liquids are actually capable of supporting transverse or shear modes.  But according to Frenkel's proposition a liquid should support transverse modes provided the frequency satisfies (frequency $\omega>\frac{1}{\tau}$, see below).


Nevertheless the idea that at least longitudinal phonons could be excited below $0.5$ K  has traditionally been taken as the explanation for the observation that the heat capacity of liquid helium varies as $T^{3}$  \cite{feynman}. In most cases investigations have been limited to narrow ranges of temperature and pressure these being of immediate concern in early experiments. Therefore, it is some interest to examine the thermodynamic properties of liquid helium both at elevated pressures, but still presenting a liquid phase, and also for wider ranges of temperature.

 
 

Accordingly in this paper we present the heat capacity of bosonic liquid helium as determined within in the framework of a phonon theory of liquids. The physical picture of elementary excitations is clarified by means of a study of phonon contributions, both longitudinal-like and transverse-like, to the energy spectrum of liquid helium at elevated pressures. From the analysis of experimental data of viscosity and heat capacity \cite{NIST} and theoretical calculations we find that liquid helium remains quantum at elevated pressures for a broad temperature range.
\begin{figure}
\begin{center}
{\scalebox{0.66}{\includegraphics{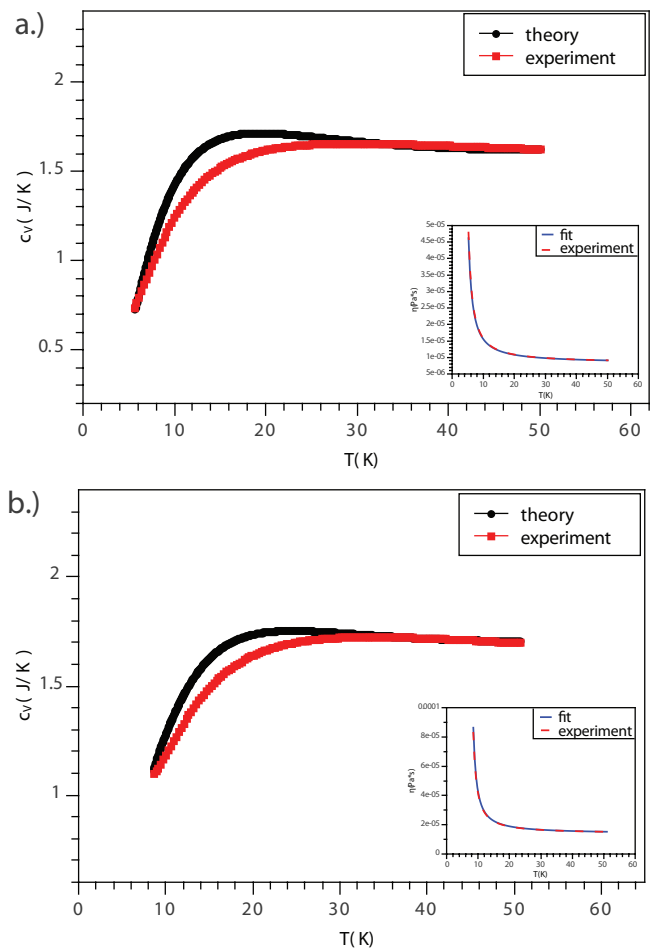}}}
\end{center}
\caption{Experimental and calculated $c_v$, experimental and interpolated $\eta$ (insets) for liquid helium. Experimental $c_{V}$ and $\eta$ are taken from the NIST database at pressures 20 MPa (Fig. \ref{fig1}.a) and 45 MPa (Fig. \ref{fig1}.b). Values of $\tau_{D}$ used in the calculation are 1.41 ps ($^{4}$He at 20 MPa) and 1.40 ps ($^{4}$He at 45 MPa). Values of $G_{\infty}$ are 0.0070 GPa and 0.0074 GPa. Experimental values of $\alpha$ calculated from the NIST database at the corresponding pressures above are $8.7\cdot 10^{-3}$ K$^{-1}$ and $8.3\cdot 10^{-3}$ K$^{-1}$. Values of $\alpha$ used in the calculation are $3\cdot 10^{-3}$ K$^{-1}$ and $2.8\cdot 10^{-3}$ K$^{-1}$. The uncertainty of both experimental heat capacities and viscosities is about 5-10\% \cite{NIST}. Insets also show viscosity fits.}
	\label{fig1}
\end{figure}

\section{The phonon theory of liquids} 
The fact that the solid-like value of heat capacity in liquids at the melting point  may be summarized by $C_{V}=3 N k_{B}$, and the fundamental observation that liquids retain the property of fluidity leaves us with an apparent contradiction. In order to reconcile it,  and as noted above, J. Frenkel introduced the average time between two consecutive atomic jumps thus providing a microscopic description of Maxwell's phenomenological visco-elastic theory of liquid flow \cite{maxwell}.  If $\tau$ is large compared with the period of atomic vibrations, a liquid is characterized by vibrational states as in a solid (solid glass) including shear modes with frequency $\omega>\frac{1}{\tau}$. If $\tau$ is small compared with time during which an external force acts on a liquid, usually liquids flow. The solid-like ability of liquids to sustain high-frequency propagating modes down to wavelengths on the atomic scale, at the temperature around and above melting point, was  observed fairly recently \cite{copley,pilgrim,burkel,morkel,grimsditch}. 

As noted, $\tau$ is a fundamental flow property of a liquid, and is directly related to liquid viscosity $\eta$: $\eta=G_{\infty}\tau$ \cite{frenkel, maxwell}, where $G_{\infty}$ is the instantaneous shear modulus. According to time scale the motion of an atom in a liquid can be viewed as of two types: quasi-harmonic vibrational motion around an equilibrium position as in a solid glass, with Debye vibration period of about $\tau_{D}=$0.1 ps, and diffusive motion between two neighboring positions, where typical diffusion distances exceed vibrational distances by about a factor of ten. When $\tau$ significantly exceeds $\tau_{D}$, the number of diffusing atoms and, therefore, the diffusing energy, becomes small and can be ignored.

The phonon theory of liquids allows us to calculate liquid internal energy in general form which can be compactly presented as
\begin{equation}\label{lenergy}
E=NT\left(1+\frac{\alpha T}{2}\right)\left(3D\left(\frac{\hbar\omega_{\rm D}}{T}\right)-\left(\frac{\omega_{\rm F}}{\omega_{\rm D}}\right)^3D\left(\frac{\hbar\omega_{\rm F}}{T}\right)\right)
\end{equation}
where $D(x)=\frac{3}{x^3}\int\limits_0^x\frac{z^3{\rm d}z}{\exp(z)-1}$ is the Debye function \cite{landau}, $\alpha$ is the thermal expansion coefficient, $N$ is the number of phonon states, $T$ is the temperature, and $\omega_{D}$ and $\omega_{F}$ are Debye and Frenkel frequencies correspondingly.

Eq.(\ref{lenergy}) accounts for longitudinal and also for two high-frequency shear modes with frequency $\omega>\frac{1}{\tau}$. It originates at the same level of approximation as Debye's phonon theory of solids by using the quadratic density of states. The result for a harmonic solid follows from Eq.(\ref{lenergy}) when $\omega_{F}=0$, corresponding to infinite relaxation time, and thermal expansion coefficient $\alpha=0$. This theory of liquids incorporates the effects of anharmonicity and thermal expansion, which is very important not only for classical liquids such as Hg and Rb \cite{trachenko}, but also for liquid helium as we can see further.

The phonon theory of liquids has recently been formulated in a form that predicts the heat capacity of 21 different liquids, among those: noble, metallic, molecular and hydrogen-bonded network liquids. The theory covers both the classical and quantum regimes and agrees with experiment over a wide range of temperatures and pressures \cite{bolmatov}. 

\section{Heat capacity of liquid helium}
Accordingly we now take a derivative of energy $E$ (Eq.(\ref{lenergy})) with respect to temperature $T$ at constant volume and compare it to experimental data of heat capacity per atom: $c_{V}=\frac{1}{N}\frac{dE}{dT}$. We have used the National Institute of Standards and Technology (NIST)
database \cite{NIST} that contains data for many chemical and physical quantities, including $c_{V}$ for liquid helium. We aimed to check our theoretical predictions in a wider range of temperature, and therefore selected the data at pressures significantly exceeding the critical temperature and pressure of liquid helium ($T_{c}=5.1953$ K and $P_{c}=0.22746$ MPa) where it exists in a liquid form in the broader temperature range.  As a result, the temperature range in which we calculate $c_{V}$ is about 40-45 K. Viscosity data was taken from the same database \cite{NIST}, and fitted in order to use in Eq.(\ref{lenergy}) to calculate $c_{V}$. We used the Vogel-Fulcher-Tammann (VFT) expression to fit the viscosity data: $\eta=\eta_{0}\exp{A/(T-T_{0})}$. To calculate $c_{V}$ from Eq.(\ref{lenergy}), we have taken viscosity data at the same pressures as $c_{V}$ and converted it to $\tau$ using the Maxwell relationship $\tau=\frac{\eta}{G_{\infty}}$, where the Frenkel frequency can be conveniently expressed as 
$\omega_{F}=\frac{2\pi}{\tau}=\frac{2\pi G_{\infty}}{\eta}$.

Eq.(\ref{lenergy}) has no fitting parameters, because the parameters $\omega_{D}$, $\alpha$ and $G_{\infty}$ are fixed by system properties. Values of these parameters used in Eq.(\ref{lenergy}) are in a good agreement with typical experimental values. There is a difference between the experimental $\alpha$ and the $\alpha$ used in the calculation. At each pressure the experimental $\alpha$ was estimated from the formula $\alpha=\frac{1}{V}\frac{\Delta V}{\Delta T}$. Experimentally, V$\propto$T only approximately. Consequently, we approximated V$=$V(T) by a linear dependence (an approximation results in somewhat different $\alpha$ used in Eq.(\ref{lenergy})). Further, $\tau_{D}=\frac{2\pi}{\omega_{D}}$ used in the calculation (see the caption in Fig. \ref{fig1}) is consistent with the known values for low temperature liquid helium under pressure that are typically in the 1-2 ps range \cite{humprey}. The uncertainty of both experimental heat capacities and viscosities is about 5-10\% \cite{NIST}.

\section{Discussions and conclusions}
As noted earlier, there are two basic analytical approaches to calculate liquid energy and heat capacity: from the gas phase and from the solid. The approach from the classical gas phase has two main contributions to liquid energy; kinetic and potential parts and can be presented as
\begin{equation}\label{genergy}
E=K+\int gUdV
\end{equation}
where K is kinetic energy, g is normalized correlation function and U is the interatomic energy. Generally the expression in Eq.(\ref{genergy}) is difficult to evaluate for a many-body systems and  it is not clear how to rigorously incorporate quantum effects at elevated temperatures (say tens of Kelvins for He) into Eq.(\ref{genergy}). To describe the behaviour of $c_{V}$ on the basis of Bose-Einstein condensation (BEC) in liquid helium at low temperatures was originally initiated by F. London \cite{london}. The heat capacity of BEC varies as $T^{3/2}$ at low temperatures and must pass through a maximum. This maximum has the character of a cusp which appears at critical temperature $T_{c}$ and it has nothing to do with a  subsidiary maximum  which liquid helium possesses at elevated pressures and higher temperatures (see Fig. \ref{fig1}).


The odd behaviour of the heat capacity of liquid helium at elevated pressures can now be explained in the framework of the phonon theory of liquids. When we calculate liquid energy and heat capacity in this theory the problem of strong interactions is avoided from outset and based on displacive physics associated with phonon contributions. We predict that transverse waves exist in liquid helium at high pressures. This
prediction can be verified in a future experimental work. The experimental data for heat capacity and viscosity of liquid helium confirms our hypothesis \cite{NIST}. Even at elevated pressures liquid helium persists at temperatures which are 3-5 times lower than Debye's temperature (see Fig.\ref{fig1}). As the temperature is raised in liquid helium,  more longitudinal and transverse-like phonons become progressively excited and therefore the heat capacity rapidly grows, which is very abnormal for ordinary liquids \cite{bolmatov}. Thus liquid helium at this P-T region persists as a quantum liquid and also as a solid-like liquid with non-static shear rigidity, similar to classical liquids. When the Debye and the Frenkel temperatures become roughly comparable, $c_{V}$ of liquid helium enters the saturation region ('hump'). Further increase of temperature then leads to the dissipation of transverse-like waves and $c_{V}$ of liquid helium becomes shallow (see Fig.\ref{fig1}), implying that at very high temperatures $c_{V}$ reaches its asymptotic value $3/2$.


The results just presented appear to be fairly accurate over the temperature range of experimental importance, despite the fact that the formal expression for liquid internal energy is quite trivial (see Eq.(\ref{lenergy})). The agreement is somewhat worse at intermediate part (the slight maximum or 'hump' region mentioned earlier) of the $c_{V}$ curve, however the maximal difference between the predicted and experimental values is actually comparable to the experimental uncertainty of $c_{V}$, namely of 0.1-0.2 J/K \cite{NIST}.  But this can also be attributed to an oversimplified form of the Debye spectrum of phonon-like states used in our analysis.


Liquid helium in its normal state and at atmospheric pressure is actually above the conditions required for a Frenkel line (see the diagram \cite{brazhkin}) and barely supports transverse-like elementary excitations. At  significant pressures (around 10 GPa) liquid helium quite resembles the other noble-gas liquids; there are transverse-like excitations but at its melting temperature almost all phonons are already excited. This state of liquid helium is rigid but not quantum. Here we are suggesting an interesting intermediate pressure range of  5~MPa-500~MPa where liquid helium near melting temperature is already solid-like but is still significantly quantum.

\section{Acknowledgements}
 D. Bolmatov thanks Myerscough Bequest and K. Trachenko thanks EPSRC for financial support.
D. B. acknowledges Thomas Young Centre for Junior Research Fellowship and Cornell University (Neil Ashcroft and Roald Hoffmann) for hospitality. We acknowledge Neil Ashcroft for fruitful discussions.

\end{document}